\documentclass{elsart}

\usepackage{graphicx}
\usepackage{amsfonts}
\usepackage{amssymb}

\newlength{\figwidth}
\begin{document}
\begin{frontmatter}
\title{Development of Shashlyk Calorimeter for KOPIO}
\author[INR,Yale]{G.S.~Atoian},
\author[INR,Yale]{V.V.~Issakov},
\author[INR]{O.V.~Karavichev},
\author[INR]{T.L.~Karavicheva},
\author[INR,Yale]{A.A.~Poblaguev\corauthref{email}}, and
\author[Yale]{M.E.~Zeller}
\address[INR]{Institute for Nuclear Research of Russian Academy of
Sciences, Moscow 117312, Russia}
\address[Yale]{Physics Department, Yale University, New Haven, CT 06511, USA}
\corauth[email]{Corresponding author. 
{\em E-mail address:} poblaguev@bnl.gov (A.A.~Poblaguev)}
\date{10 October 2003}
\begin{abstract}

A Shashlyk calorimeter prototype for the KOPIO experiment has been constructed and experimentally tested. The energy resolution of about $4\%/\sqrt{E\mathrm{(GeV)}}$ for $0.5\div2.0~\mathrm{GeV/c}$ positrons was obtained.
Based on this results as well as on the results of special measurements, a Monte-Carlo model of the Shashlyk module response was developed. This model, including the effects of shower evolution, light collection in scintillator plates, light attenuation in fibers, quantum efficiency of the photodetector, thresholds and noises in the readout system is consistent with experimental results.
Possible improvment of the Shashlyk energy resolution up to $3\%/\sqrt{E\mathrm{(GeV)}}$, the level required by KOPIO experiment, are discussed.

\end{abstract}
\begin{keyword}
Shashlyk calorimeter \sep Monte-Carlo simulation
\PACS 29.40.Vj \sep 07.05.Tp
\end{keyword}

\end{frontmatter}

\section{Introduction.}

	The KOPIO experiment \cite{KOPIO} at the Brookhaven National Laboratory (BNL) Alternating Gradient Sinchrotron (AGS) is designed to measure the decay rate for
$K_L^0\to\pi^0\nu\bar{\nu}$, a ``gold plated'' CP-violating process \cite{Littenberg}. 
This experiment will provide the cleanest determination of the fundamental parameter 
 that quantifies the phenomenon of CP violation in the context of the Standard Model. A measured decay rate very different from the precise expectation of the Standard Model, or one in conflict with CP violation results from the B sector, would be evidence for a new physics process.

A Photon Calorimeter which will occupy an area $5.3\times5.3~\mathrm{m}^2$ is one of the keystone elements of the KOPIO detector. Studies of Detector optimization led to the following requirements for photon detection, in the energy range of $100\div500~\mathrm{MeV}$:
\begin{itemize}
\item Energy resolution $(3\div3.5)\%/\sqrt{E\mathrm{(GeV)}}$
\item Time resolution about $80~\mathrm{ps}/\sqrt{E\mathrm{(GeV)}}$
\item Calorimeter granularity $\sim10\times10~\mathrm{cm}^2$
\end{itemize}

Shashlyk based calorimeter meets the specified requirements in an economical way. Such a calorimeter is composed of Shashlyk modules, which are lead-scintillator sandwiches read out by means of Wave Length Shifting
(WLS) fibers passing through the holes in scintillator and lead. 

     The first Shashlyk calorimeter was designed and manufactured at
Institute for Nuclear Research (Moscow) \cite{NIM} in 1991
for the
experiment 865 \cite{E865}
(Search for the Lepton Number
Violating Decay $K^+\to\pi^+\mu^+ e^-$) at the BNL AGS.
During the five-year high intensity run of the experiment, the 
Shashlyk calorimeter was a very stable and
reliable detector.
Its features, together with its low cost and well understood
method of construction, make this type of calorimeter
a good candidate for other experimental projects.  Similar
calorimeters were  built later for PHENIX experiment at RHIC (BNL)
\cite{PHENIX} and for HERA-B experiment at DESY \cite{HERA-B}.
A Shashlyk
calorimeters was also studied as a candidate for the
CMS experiment at LHC (CERN) \cite{CMS}, and one is now under construction
for LHCb experiment \cite{LHCb}.

     The E865 and other constructed calorimeters were designed to have
an energy resolution about $8\%/\sqrt{E\mathrm{(GeV)}}$.  Significant improvements
in calorimeter module construction can be made it possible to achieve 
the resolution $\sim 3\%/\sqrt{E\mathrm{(GeV)}}$ required by experiment KOPIO.

The purpose of this paper is to study the ways
of such an upgrade. 
     Based on experience with the E865 module, a new prototype module
has been designed, constructed and tested. 
As will be shown below this module provides an energy
resolution about $4\%/\sqrt{E\mathrm{(GeV)}}$.  While this improved resolution is not 
sufficient to meet the requirements of KOPIO, experimental studies of 
this module provide a reference point for tuning Monte-Carlo simulations. 
From such simulations, recommendations for constructing a Shashlyk calorimeter 
for KOPIO are made. 

\section{Design of KOPIO prototype module.}

\begin{figure}
\mbox{\includegraphics*[width=\figwidth]{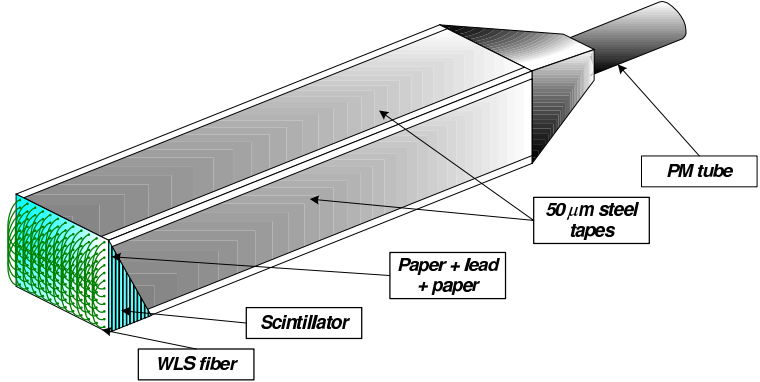}}
\caption{Shashlyk module design.}
\label{FIG01}
\end{figure}

     The design of a  prototype module for KOPIO 
is shown in Fig. \ref{FIG01}. 
     Eighteen such modules had been produced at TECHNOPLAST 
(Vladimir, Russia).

\subsection{Mechanical construction  of module}

The module is a sandwich of alternating
perforated stamped lead and injection molded polystyrene-based scintillator
plates. The transverse size of module is $110\times110~\rm{mm}^2$; the 
thicknesses of plates are $0.35\ \rm{mm}$ for lead and $1.5\ \rm{mm}$ for 
scintillator.
Each plate has 144 holes equidistantly arranged as $12\times12$ matrix,
with the spacing between the holes being $9.5~\rm{mm}$. The diameters of the 
holes are 1.5 mm in the lead plates, while the holes in the 
scintillator  have a conical shape with diameter ranging from 1.4 to 1.5 mm. 
72 WLS fibers are inserted into these holes. Each
fiber is looped at the front of the module, so that both ends of a fiber are
viewed by the photo-multiplier tube (PMT). Less than $10\%$ of light is lost while passing the loop, which radius is about $3\ \mathrm{cm}$.
The fiber ends are collected in one
bunch, cut and polished, and connected to a  PMT.
To improve the light collection $60\ \mathrm{\mu m}$ perforated white reflecting  
paper (TYVEK) is interleaved between lead and scintillator plates, and edges
of scintillator plates are aluminized.
The complete stack of all plates is held in
compression  by the four $50\ \mu$m stainless steel side strips that are
pretensioned and welded to both front and back special endcaps.

     Parameters of module are summarized in Table \ref{Tab1}.

\begin{table}
\caption{Parameters of the prototype Shashlyk module for KOPIO.}
\label{Tab1}
\begin{tabular}{ll}
\hline
Transverse size                     & $110\times110~\mathrm{mm}^2$ \\
Number of the layers                & 240                          \\
Polystyrene scintillator thickness  & $1.5~\mathrm{mm}$ \\
Lead absorber thickness             & $0.35~\mathrm{mm}$          \\
Reflective paper (TYVEK) thickness  & $2\times0.06~\mathrm{mm}$ \\
Number of holes per layer           & $12\times12$               \\
Holes spacing                       & $9.5~\mathrm{mm}$                     \\
Holes diameter in Scintillator/Lead & $1.4/1.5~\mathrm{mm}$                  \\
WLS fibers per module               & $72\times1.3~\mathrm{m}\approx92~\mathrm{m}$\\
Diameter of WLS fiber               & $1.0~\mathrm{mm}$, ($1.2~\mathrm{mm})$ \\
Diameter of fiber bundle            & $14~\mathrm{mm}$, ($17~\mathrm{mm}$) \\
Effective radiation length $X_0$    & $31.5~\mathrm{mm}$                     \\
Effective Moli\`{e}re radius $R_M$  & $54.9~\mathrm{mm}$	          \\
Effective density                   & $2.75~\mathrm{g/cm^3}$              \\
Active length                       & $473~\mathrm{mm}$ ($15.9 X_0$)      \\
Total length (without photodetector)& $610~\mathrm{mm}$                   \\
Total weight                        & $18.0~\mathrm{kg}$                  \\
\hline
\end{tabular}
\end{table}

\subsection{WLS fibers and PM tubes}

We have used three different types of fibers (KURARAY, $1~\mathrm{mm}$ diameter 
Y11(200)M-DC, and BICRON, $1~\mathrm{mm}$ diameter BCF-99-29A-SC and 
$1.2~\mathrm{mm}$ diameter BCF-92-SC) and  different 
types of PMT's (Russian FEU85 and FEU115, and 9903B of Electron Tubes 
Ltd (9903B)) for the tests. 

\begin{figure}
\mbox{\includegraphics*[bb=0 15 270 270,width=\figwidth]{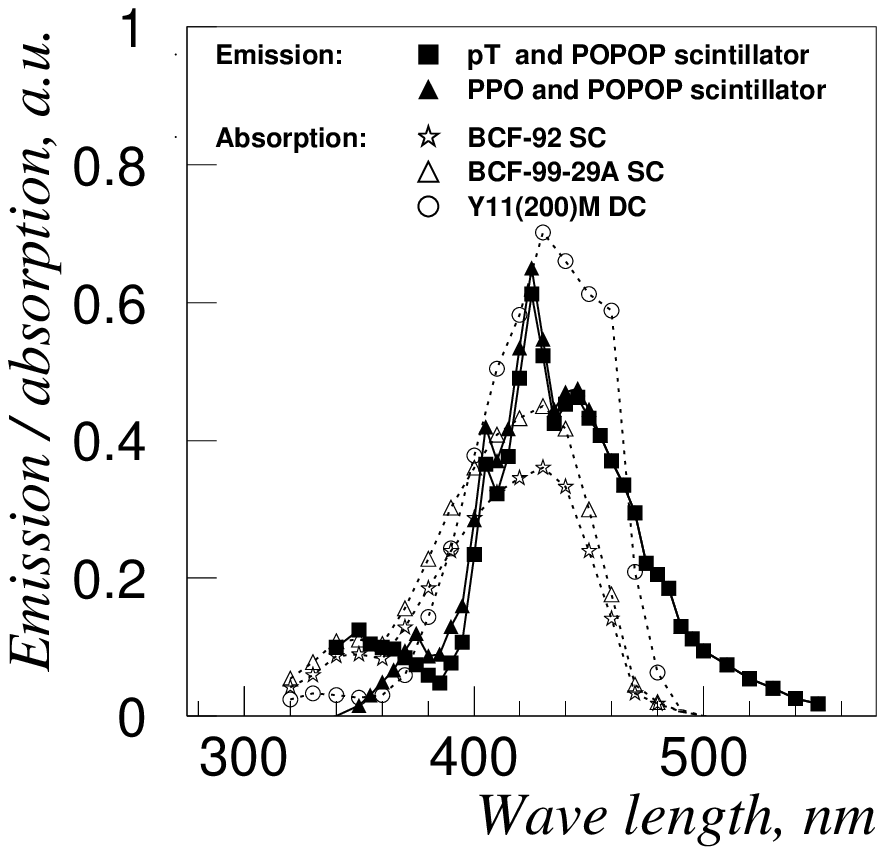}}
\caption{Absorption 
spectra of WLS fibers in comparison with emission spectra of scintillators.}
\label{FIG02}
\end{figure}

     The properties of WLS fibers used in our measurements are summarized in
Table \ref{Tab2}. These data were taken from manufacturer's Catalogs.
Experimental measurements \cite{Semenov_priv} of the absorption spectra of WLS
fibers in comparison with emission spectra of the scintillators are shown
in Fig. \ref{FIG02}.

\begin{table}
\caption{Properties of WLS fibers. PS stands for polystyrene, PMMA for polymethylmetacrylate, and FP for fluorinated polymer.}
\label{Tab2}
\begin{tabular}{lccc}
\hline
WLS fiber      & BCF-92 & BCF-99-29A & Y11(200)M-DC \\
\hline
Manufacturer   & BICRON & BICRON     & KURARAY      \\
\hline
WLS fluor      & G2, $100~\mathrm{mg/l}$ & G2, $200~\mathrm{mg/l}$ 
                                         & K27, $200~\mathrm{mg/l}$ \\
\hline
Emission peak, nm & 492 & 492 & 476 \\
\hline
Absorption peak, nm & 410 & 410 & 430 \\
\hline
Decay time, ns      & 2.7 & 2.7 & $\sim$7 \\
\hline
Core material,  & PS & PS & PS \\
refractive index & $1.59$      & $1.59$       & $1.59$       \\
\hline
Inner cladding,   & PMMA         & PMMA           & PMMA   \\
refractive index  & $1.49$       & $1.49$         & $1.49$       \\
\hline
 Outer cladding,        &   ---          &        ---       &  FP    \\
refractive index  &                &                  & $1.42$       \\
\hline
Trapping efficiency, \% & 3.1 & 3.1 & 5.4 \\
\hline
Attenuation length, m & $\sim3.5$  & $\sim4.2$ & $>3.5$                  \\
(for long fiber) & & &                                 \\
\hline
Fiber diameter, mm & 1.2  & 1.0 &  1.0                                   \\
\hline
\end{tabular} 
\end{table}

     FEU85 is an eleven stages green-extended one inch PM tube with a good quantum
efficiency for green light (Q.E. $\sim$ 15-20 \%). The maximum anode current of
this PMT is limited to $\sim5$ mA per pulse, therefore it was used with a fast
preamplifier with programmable gain K = $2\div10$.
 
FEU115 is an inexpensive
high gain eleven stages PMT with an S20 photo-cathode response and 
with typical Q.E. $\sim12\%$ for green light.

9903B  is a ten
stages green-extended tube with Rubidium bialkali (RbCs) photo-cathode 
and BeCu dynodes for lower gain $(0.2\times10^6)$,
but with extended linearity (better than 1\% up to 50 mA ) and a high short 
and long term gain stability. 9903B has a typical Q.E. $\sim18\%$ for 
$500\ \rm{nm}$. 

\subsection{Light collection}

     Electrons and positrons from an electromagnetic shower
produce ultraviolet and blue light in the scintillator plates which
is absorbed and reemitted as green light in the WLS fiber.
About $10\%$ of the green light is captured and transmitted by the fiber.

Due to the Shashlyk design, 
only light satisfying the criteria of a total internal reflection
within the scintillator plate is captured by fibers.
Since scintillator plates are thin compared to the average distance of light
to the fibers, many reflections from the surface of scintillator
occur.  Thus, high reflection efficiency is very important for good performance.
An internal reflection efficiency about $93\%$ is achievable for realistic 
surfaces.  It should also be noted that
modern technologies allow comparable diffusion reflection
efficiencies to be realized by special chemical 
treatment of the surfaces \cite{Semenov}.

Light collection uniformity in the direction transverse to the
length of the module is also important in achieving good energy resolution. 
This requires good reflectivity at the edges of the scintillator plates
since the total internal reflection at these surfaces is not necessary condition for the possibility of light penetrating to fiber. Thus the scintillator 
edges were aluminized to increase light reflection.

\section{Experimental study of prototype module}

     The characteristics of the modules were studied on B2 test beam
at AGS with $0.5\div2\ \mathrm{GeV/c}$ positrons and pions. 
Measurements were done during the Spring and Fall test runs 1998. The prototype
of calorimeter ($3\times3$ modules) was mounted on a platform which could 
be moved
horizontally and vertically with respect to the beam line. 
Upstream of the calorimeter, a trigger counter telescope
was installed which consisted of three beam defining scintillators (S1 - S3)
and a $2X_0$ veto lead-scintillator counter with a $15\ \rm{mm}$ 
diameter hole through which the beam passed.   A Cerenkov counter was used for
identification of positrons. 

     The signals from PM tubes were read out using an $80\ \rm{ns}$ wide gate 
with a LeCroy 1885 FASTBUS ADC providing 12 bits of dynamic range. A
separate channel, including a PM tube without a Shashlyk module and
located inside the detector box, was used to check for channel to channel
coherent noise correlations. It was often found that the coherent noise was 
very high. A possible explanation of this effect is that there was no
``clean ground'' in the  test beam area. The data from the runs with increased 
coherent noise were skipped in the off-line analysis.  

     Three types of triggers were used during the beam tests for monitoring and
data taking : a random trigger to monitor pedestal behavior and to check noise
correlations, and two beam triggers (with and without the veto counter) for
measurements. 

     Each module was individually calibrated using either the minimum
ionization peak from high momentum pions, or the deposited energy peak from
$1\ \mathrm{GeV}$ positrons. In each calibration  measurement about 5000 
particles  were passed through the central region of each module at normal 
incidence.  Using these data, the calibration
coefficients were estimated with precision better than $1\%$.

     The energy resolution was determined only for positrons passing through  
the $1\times1\ \rm{cm}^2$ area in the center calorimeter as defined by 
S3 scintillator.

     The momentum spread of the beam significantly  contributed to the 
apparent energy resolution in the modules at all measured energies. 
For beam momentums above $1~\mathrm{GeV/c}$, the spread was about $1\%$.  However, according
to a GEANT \cite{GEANT} calculation, $0.5\ \mathrm{GeV}$ positrons passing 
through
matter (15 m of Air and 3.5 cm of scintillation counters)
lose about $15\ \mathrm{MeV}$ of their energy with fluctuations of about 
$\sigma\sim 10\ \mathrm{MeV}$. This energy loss required significant
corrections to the results of the measurements.

     The intensity of beam was chosen sufficiently low to reduce pile-up
and rate effects during the spill, but due to the large beam size this 
requirement was only partially met.  The beam particle rates for a nonet of 
modules were ($50\div100$)K per spill.  Small corrections for pile-up  and rate 
effects
were taken into account in data analysis. 

     The electronic noise term for the nonet of modules was measured during
special test when calorimeter was removed from the beam. Its contribution
to the energy resolution of the nonet is $5\pm1\ \mathrm{MeV}$ for FEU85 PM tubes 
(with preamplifiers) and it is $2\pm1\ \mathrm{MeV}$ for other PM's 
(without preamplifiers). 

     The sum of electronic noise and pile-up effect for the nonet was
measured in other special beam test with the trigger gate for the ADC shifted 
by about
$300\ \rm{ns}$. The contribution of this effective noise to the energy 
resolution was $10.3\pm1.3\ \mathrm{MeV}$ for FEU115 PM tubes. 
This value is somewhat larger than expected from pile-up effect, probably 
due to contributions of ``after pulse noise'' typical 
for this type of PM tube. The total  equivalent noise for FEU85 PM
 tubes
with preamplifiers was $6.5\pm1\ \mathrm{MeV}$. The lowest equivalent noise, 
$3.6\pm1\ \mathrm{MeV}$, was obtained with 9903B. 

\begin{figure}
\mbox{\includegraphics*[width=\figwidth]{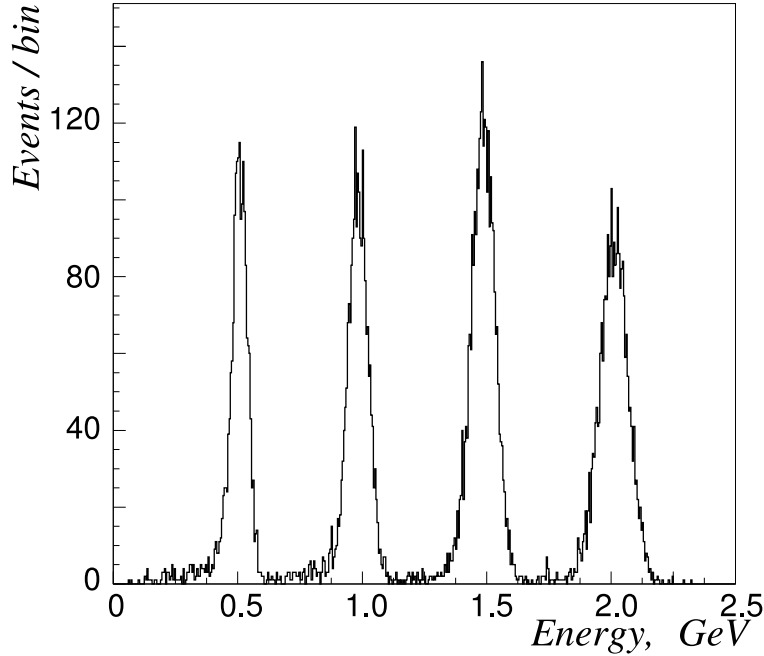}}
\caption{The typical pulse height spectrum for the nonet of Shashlyk
        modules for $0.5\div2.0\ \mathrm{GeV/c}$ positrons.} 
\label{FIG03}
\end{figure}

 Signal distributions for 0.5, 1.0, 1.5 and 
2.0 $\mathrm{GeV/c}$ positrons are shown in Fig. \ref{FIG03}. 
Only modules with signals above threshold
$\sim3\ \mathrm{MeV}$ were included to the sum over nonet.
We believe that low energy tails for the low momentum 
(0.5 and 1.0 $\mathrm{GeV/c}$) positrons are due to the interaction of the positrons
with matter upstream of the calorimeter modules.

\begin{table}
\caption{Results of experimental study of Shashlyk prototype (240 layers 
of 0.35 mm lead and 1.5 mm scintillator) with 1 mm diameter 
KURARAY Y11(200)M-DC fibers. Energy resolution is corrected for beam
momentum spread and energy loss upstream of the calorimeter.}
\label{Y11_data}
\begin{tabular}{cccccc}
\hline
$p_e$ & N      & $\langle A_e \rangle$ & $\sigma_A$ & Measured         & Corrected        \\
$\mathrm{GeV/c}$ & events & ADC cnts              & ADC cnts   & $\sigma_E/E$(\%) & $\sigma_E/E$(\%) \\
\hline
2.00 & 3000 & 1430 & 42   & 2.92$\pm$0.05   & 2.70$\pm$0.09   \\
1.75 & 3000 & 1241 & 39   & 3.14$\pm$0.05   & 2.91$\pm$0.09   \\
1.50 & 3000 & 1056 & 35   & 3.30$\pm$0.05   & 3.06$\pm$0.10   \\
1.25 & 3000 &  913 & 34   & 3.68$\pm$0.07   & 3.43$\pm$0.11   \\
1.00 & 3000 &  699 & 29   & 4.20$\pm$0.07   & 3.91$\pm$0.11   \\
0.75 & 3000 &  531 & 25   & 4.77$\pm$0.09   & 4.45$\pm$0.13   \\
0.50 & 2500 &  359 & 23   & 6.29$\pm$0.17   & 5.56$\pm$0.20   \\
\hline
\end{tabular}
\end{table}

     The experimental results for energy resolutions for the calorimeter 
prototype with
KURARAY Y11(200)-M-DC fibers are shown in 
Table
\ref{Y11_data}. Corrections to energy resolution were obtained by 
subtracting the contributions of the beam positron momentum spread 
and the effects of energy loss upstream of the calorimeter, studied 
with GEANT Monte-Carlo, from measured values. 
 
The corrected dependence of the energy resolution on the positron energy may be 
approximated by the following function
$$ \frac{\sigma_E}{E} = (-0.1\pm0.8)\% \oplus 
                        \frac{(3.8\pm0.1)\%}{\sqrt{E}} \oplus 
                        \frac{(-0.8\pm0.6)\%}{E};~~~~0.5<E<2.0, $$
where $E$ is measured in $\mathrm{GeV}$ and $\oplus$ stands for qudratic summation.
It should be emphasized that this dependence was obtained for the nonet of modules with a narrow,
$1\times1~\mathrm{cm}^2$, beam in the center.  

\begin{figure}
\mbox{\includegraphics*[bb=0 15 270 270,width=\figwidth]{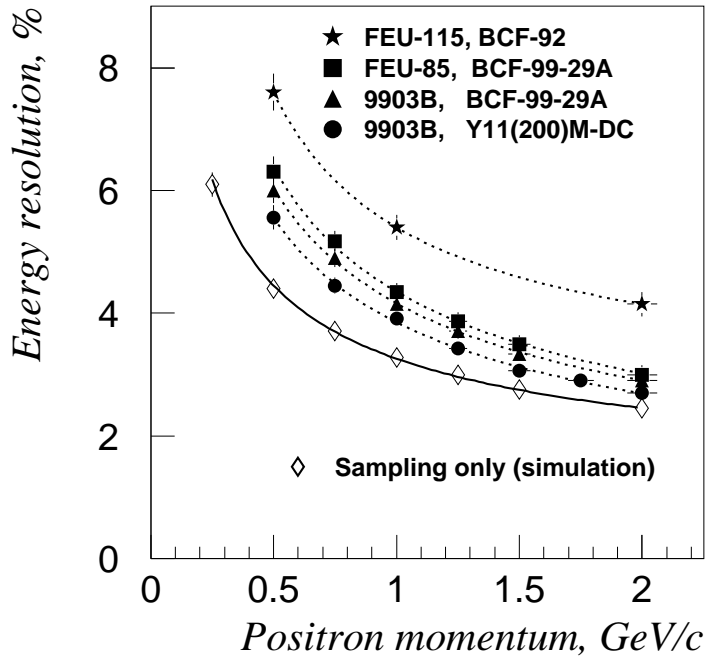}}
\caption{The energy resolution of Shashlyk calorimeter (nonet of modules)
        versus the momentum of the positron beam. The structure of calorimeter
        module is 0.35 mm lead, 1.5 mm scintillator, 240 layers. The measurements
        were done for four types of light collection system (see text).}
\label{FIG04}
\end{figure}

The measured dependence of the energy resolution on energy after correction for the positron momentum 
spread  for all studied fiber/PM tubes combinations is  shown in Fig. \ref{FIG04}. 
The GEANT calculated limit of the energy resolution due to the sampling
only is  
shown for comparison.

\section{Monte-Carlo simulation}

     Energy resolution of a Shashlyk module depends on variety
of factors. We include the following:
\begin{itemize}
\item Sampling, {i.e.} thicknesses of lead and scintillator
      plates.
\item Longitudinal leakage {i.e.} fluctuation of energy leakage
      due to the finite length of module.
\item Transverse leakage {i.e.} fluctuation of energy leakage due to the
      limited number of modules used to reconstruct an electromagnetic
      shower.
\item Effect of the presence of holes, fibers, and steel strips.
\item Light attenuation in the fiber.
\item Photostatistics.
\item Uniformity of light collection.
\item  Electronic noise.

\end{itemize}

\subsection{GEANT simulation}

\begin{figure}
\mbox{\includegraphics*[bb=0 15 270 270,width=\figwidth]{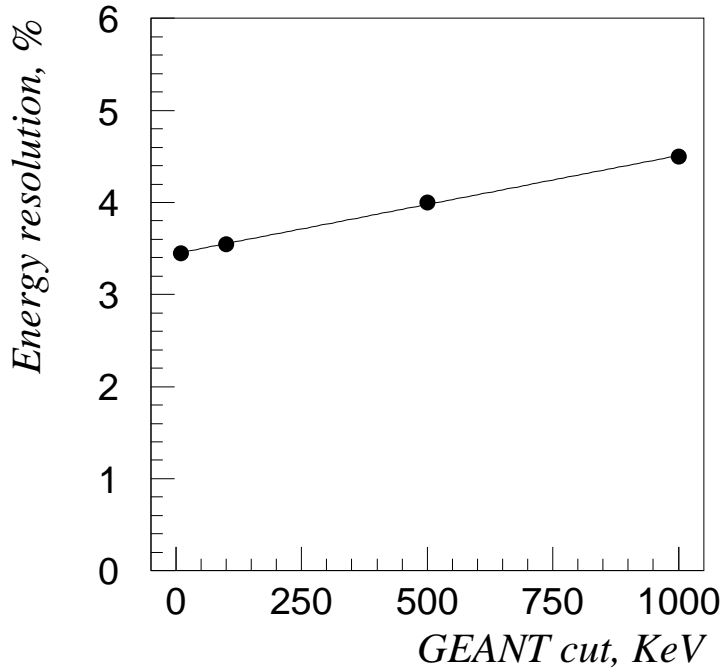}}
\caption{Energy resolution
        versus GEANT parameter cuts - \textsf{CUTELE, CUTGAM}. GEANT simulation
        was performed for the nonet of Shashlyk modules. The structure of a 
module is 0.35 mm lead, 1.5 mm scintillator, 240 layers.}
\label{FIG05}
\end{figure}

     GEANT 3.21 was used to simulate the development of electromagnetic
showers in the module.  GEANT contains many tuning parameters
which allows one to select between speed and quality of simulation.
Cuts on energy of electrons \textsf{(CUTELE)} and photons \textsf{(CUTGAM)}
are crucial for correct simulation of the response of Shashlyk modules.
Default values for both parameters are $1\ \mathrm{MeV}$.  The dependence
of the energy resolution
on the choice of these cuts is displayed in Fig. \ref{FIG05}.

\begin{figure}
\mbox{\includegraphics*[width=\figwidth]{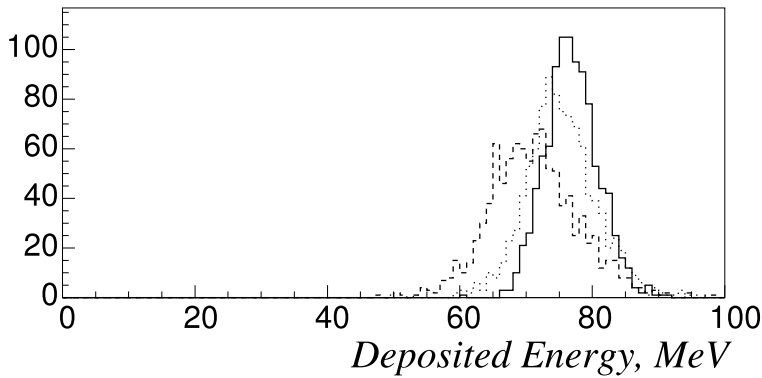}}
\caption{ GEANT simulation of the Shashlyk calorimeter response 
(energy deposited in scintillator) on $250\ \mathrm{MeV}$ photon. 
Dashed line for default cuts
        {\sf CUTELE=CUTGAM=1 MeV, DRAY=0, LOSS=2},
        dotted line for
        {\sf CUTELE=CUTGAM=10 KeV, DRAY=0, LOSS=2},
        and solid line for
        {\sf CUTELE=CUTGAM=DCUTE=10 KeV, DRAY=1, LOSS=1.}
     Calculations have been performed for long (6000 layers)
        module consisting of 0.35 mm of lead and 1.5 mm of scintillators.
     Statistics is the same for all three distributions.}
\label{FIG06}
\end{figure}

     Simulated energy resolution is also strongly dependent
details of simulating energy loss.  Default GEANT parameterization
does not include generation of delta-rays \textsf{(DRAY=0)} and uses
Landau-Vavilov-Gauss fluctuations \textsf{(LOSS=2)}
for energy loss.
A significantly different result for energy resolution is 
obtained (Fig. \ref{FIG06})
if one enables delta-ray generation \textsf{(DRAY=1)} with appropriate
modification to energy loss \textsf{(LOSS=1)}.

     Our calculations use cuts  
{\sf CUTELE=CUTGAM=DCUTE=10 KeV} 
and  generating of
delta-rays ({\sf DRAY=1, LOSS=1}) since this choice provides 
better coincidence with experimental data.

\begin{figure}
\mbox{\includegraphics*[width=\figwidth]{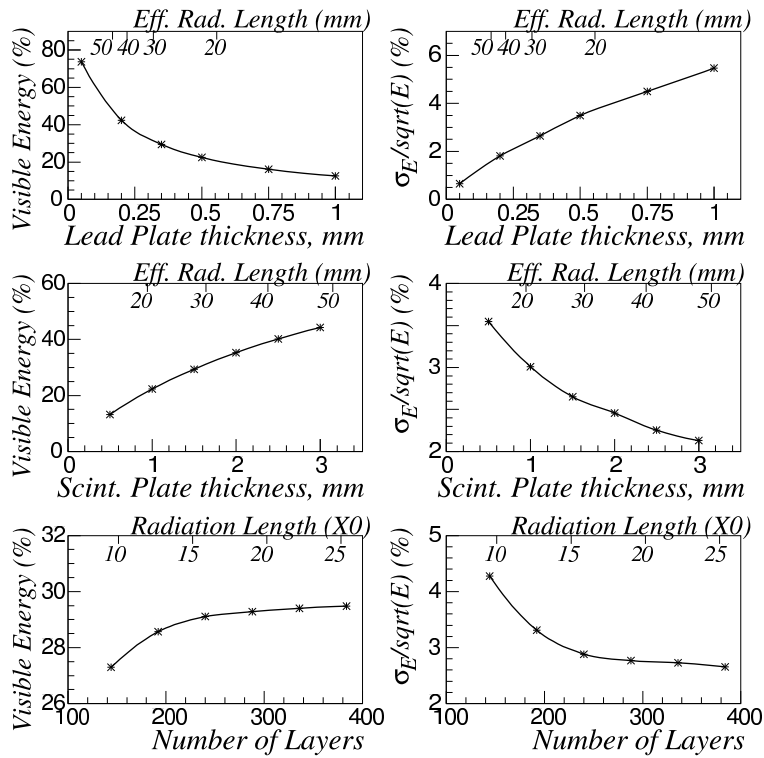}}
\caption{ Calorimeter response and energy resolution as a function
        of lead and scintillator plate thickness and number of layers.
     It is assumed that calorimeter consists of 6000 layers of 0.35 mm lead
        and 1.5 mm of scintillator unless otherwise noted in histograms.}
\label{FIG07}
\end{figure}

     The dependence of the calorimeter response (visible energy) and
energy resolution on thickness of lead and scintillator plates,
and number of layers, are shown in Fig. \ref{FIG07}. Calculations
have been performed for $250\ \mathrm{MeV}$ photons.  Unless otherwise noted 
on the histograms, the
calorimeter is modeled to be very long (6000 layers), consisting of 
$0.35\ \rm{mm}$ lead and $1.5\ \rm{mm}$ scintillator plates. 

Sampling contribution to energy resolution may be estimated by the expression
$\sigma_E/\sqrt{E}=4.7\%t^{-1/2}s^{1/4}$, where $t$ and $s$ are thicknesses
(mm) of lead and scintillator plates, respectively.
 About 20 radiation lengths
are required in order to achieve a resolution better than $3\%/\sqrt{E\mathrm{(GeV)}}$,
if only the mechanical structure of module is taken into account.

The contribution to the energy resolution of holes in the calorimeter and transverse leakage
also has been estimated in the GEANT Monte-Carlo and will be discussed below.

\subsection{Light attenuation in fibers}

\begin{figure}
\mbox{\includegraphics*[bb=0 15 270 270,width=\figwidth]{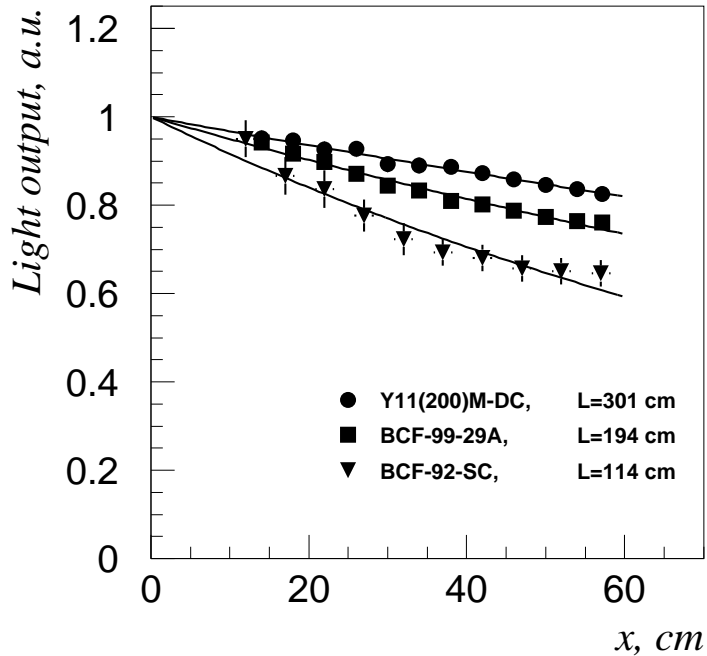}}
\caption{ The effective attenuation of the light in the fibers of Shashlyk 
module. Experimental data (marks) are fit by the exponential dependence $\exp{(-x/L)}$ 
(solid lines), where $x$ is the distance to the PMT and $L$ is the effective attenuation length.
Measurements have been performed with a muon beam. 
}
\label{FIG08}        
\end{figure}

A main concern about attenuation length in fibers is related
to the longitudinal fluctuations of electromagnetic showers.
The typical value for such fluctuations is about one radiation length,
which is 3-4 cm for modules of interest (see effective radiation length in
Fig. \ref{FIG07}).
This implies that the effective attenuation
length in fibers installed into the calorimeter modules must be greater than
$2\div3\ \mathrm{m}$ to have this contribution to energy resolution be much 
smaller than the sampling contribution.

We have experimentally measured the light attenuation in three
different fibers: (i) BCF-92-SC, (ii) BCF-99-29A, and (iii)
KURARAY Y11(200)M-DC.  Measurements were performed using
muons transversely penetrating modules.  The size of the
beam spot was $1\times1\ \rm{cm}^2$.   Results of the measurements
are presented in Fig. \ref{FIG08}. It should be noted that these 
measurements include effects of the fiber loop and 
the short distance component of light attenuation in fiber, 
{i.e.}, the effective attenuation length
in a Shashlyk module for selected PM tubes was measured.

The attenuation length in KURARAY fibers satisfies our requirements.
This fiber also provides the best light output. For Monte Carlo calculations an
attenuation length of $300\ \mathrm{cm}$ was used, a value which includes
the effect of fiber loops.

\subsection{Photostatistics}

\begin{figure}
\mbox{\includegraphics*[bb=0 15 270 270,width=\figwidth]{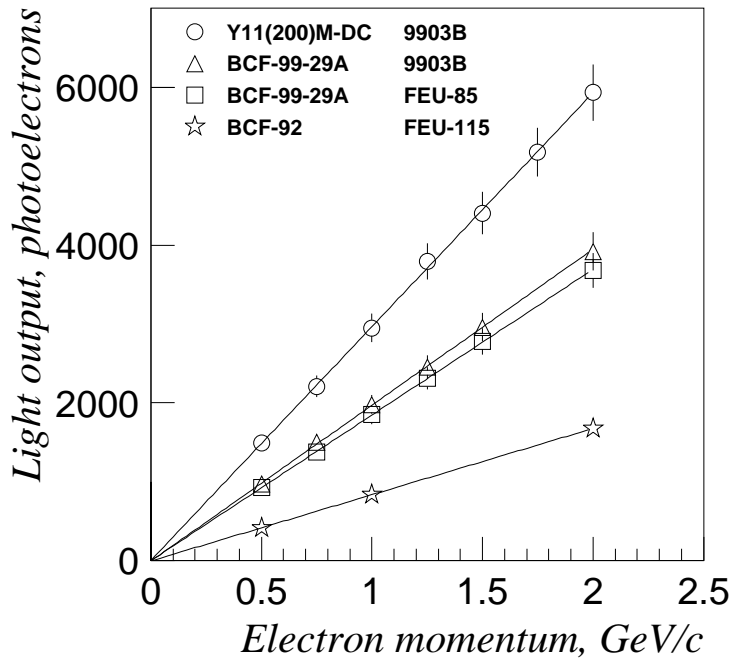}}
\caption{The light yield of the nonet of modules versus the
        momentum of the positron beam. The module structure
        is 0.35 mm lead, 1.5 mm scintillator, 240 layers. The measurements were done
        for four types of light collection systems (see text). The solid
        lines show the linearity response of detected signals for
        different energy deposition in Shashlyk calorimeter.} 
\label{FIG09}
\end{figure}

Light output was experimentally studied for several types of fibers
and photomultiplier tubes. The results are displayed in Fig. \ref{FIG09}.
The best result was obtained for KURARAY fiber Y11(200)M-DC. The light output
for a module with this fiber was experimentally found to be 3100 
photoelectrons per $\mathrm{GeV}$. This value is matched well by an estimation based
on following assumptions: (i) 7500 scintillation photons per $\mathrm{MeV}$
 of deposited energy (for molding scintillator used in prototype), 
(ii) $10\%$ efficiency of light collection from scintillator to fiber, 
(iii) $2\times5.4\%$ trapping 
efficiency in fiber, where factor 2 corresponds to the two directions of light transmission, (iv) attenuation in fiber $\sim 0.85$, (v) quantum
efficiency of 9903B photo-cathode $\sim 18\%$

\subsection{Simulation of light collection}

Uniformity of light collection in the scintillator plates is another important
ingredient for obtaining good energy resolution. 

To study the dependence of light collection on optical
parameters of scintillator plates, a special optical model
was developed. Charged particles passing through a
scintillator plate generates a number of randomly directed photons
proportional to energy deposited.  Photons
so generated may be absorbed on, reflect from,
or penetrate through the surfaces of the plate.
Three types of outer surfaces are considered: (i) ``clean'', {i.e.},
the photon is specularly reflected if the incident angle is appropriate for
total internal reflection, otherwise it exits from the scintillator,
(ii) ``mirror'', {i.e.}, the photon is specularly reflected independently of
incident angle, and (iii) ``paint'', {i.e.}, the photon is reflected 
diffusely.
Photons leaving the scintillator may enter a fiber or be reflected from
environmental ``paper'' (diffusion reflection) or ``mylar'' (mirror
reflection). Photons entering a fiber may be reemitted or may
exit the fiber, depending on the actual length of the photon track
in the fiber. The model is customized by such parameters as refractive indices
of scintillator and fiber, attenuation length in scintillator,
probability of photon reemitting from a fiber, reflection efficiencies,
and by geometry (size of the plate, hole diameter, fiber diameter).

In spite of the simplicity of this model ({e.g.}, it does not take into 
account
the dependence of reflection probability on incident angle
and polarization of photons) it gives a reasonable description of light
collection in scintillator and allows one to compare different choices
of the optics of scintillator plate. Results of light collection simulation
strongly depends on reflection efficiencies, which 
are usually poorly known since they depend on
the quality of surface treatment.  For this reason such a
model should be used for relative predictions rather than for absolute.

\begin{figure}
\mbox{\includegraphics*[bb=0 15 270 270,width=\figwidth]{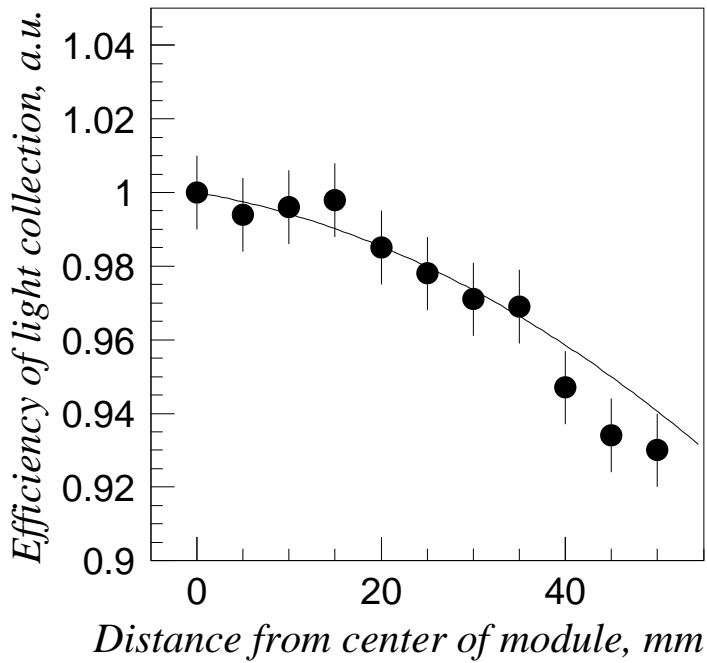}}
\caption{The nonuniformity of light collection versus
        the position of the light source. The distance is measured from the
        center of module.  The source is a is 1 cm x 1cm muon beam. The stars 
     are experimental data, the curve
        shows the optical model simulation.}
\label{FIG10}
\end{figure}

\begin{figure}
\mbox{\includegraphics*[bb=0 15 270 270,width=\figwidth]{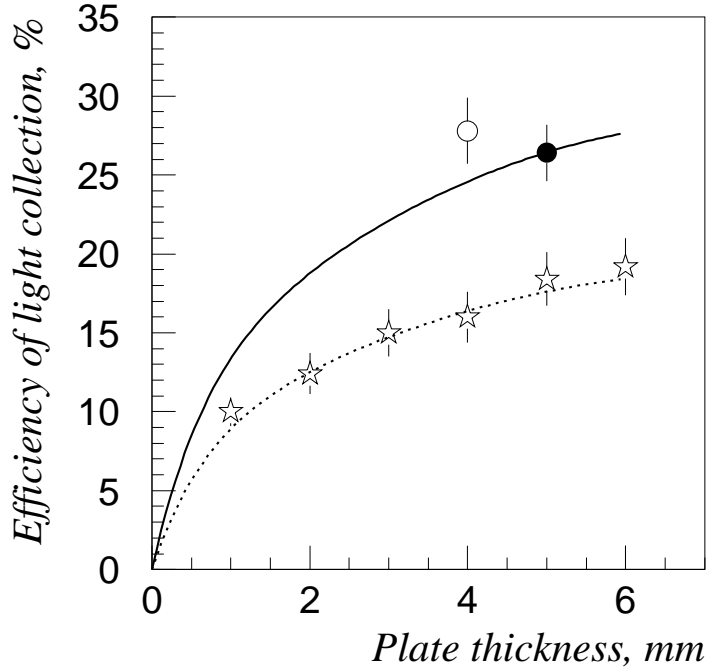}}
\caption{The collection efficiency of scintillation light from scintillator
        plate
        for green light in WLS fibers versus the thickness of scintillator plate. 
        The circles are 
        for experimental data with optical contact between fibers and 
        scintillator, the stars are for experimental data without optical 
        contact, and
        the curves show the optics model simulation.
}
\label{FIG11}
\end{figure}

\begin{figure}
\mbox{\includegraphics*[bb=0 15 270 270,width=\figwidth]{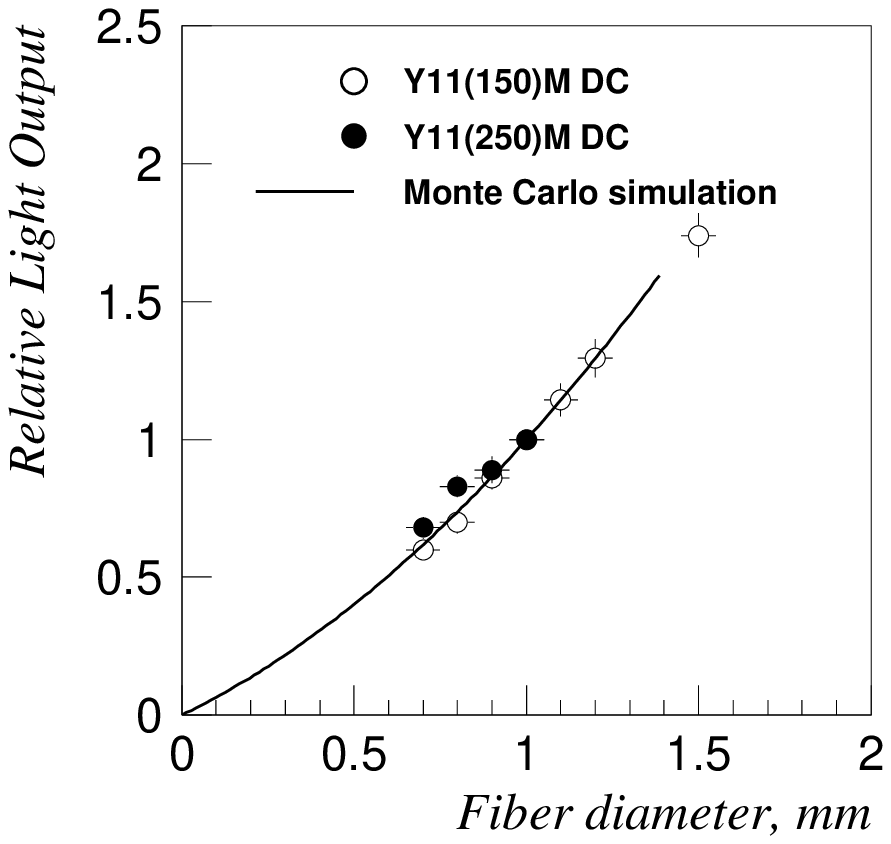}}
\caption{Relative light output for different diameter KURARAY Y11 fibers
            compared to a $1.0~\mathrm{mm}$ diameter fiber. Experimental 
            points are the data from \cite{Semenov}.
            The solid line is an optical model calculation.}
\label{FIG12}
\end{figure}

Model predictions, after tuning of all parameters,  for the dependence of 
light output on position 
of the light source, on the thickness
of scintillator plate, and on the fiber diameter are compared with
experimental measurements in Fig. \ref{FIG10}-\ref{FIG12}.

The following parameters have been used for describing the light collection
of a prototype module: total internal reflection efficiency equals $93\%$,
the large surface of scintillator plate is covered by paper with a reflection 
efficiency of $70\%$,  edges of the scintillator plates are mirrored with a 
reflection coefficient of $70\%$, absorption length in scintillator equals 
$100\ \rm{cm}$,
absorption length for scintillator light in fiber is $0.5\ \rm{mm}$,
$5\%$ of light is lost in each passing through the boundary between scintillator and fiber.
There is an air gap between fibers
and scintillator. 140 photons per $1\ \mathrm{MeV}$ of deposited
energy in scintillator were generated to match
the number of photoelectrons produced by  $1\ \mathrm{GeV}$ positron
(if only losses due to the light collection and absorption in fiber are
taken into account).

\section{Monte Carlo study of prototype module}

     To check the validity of our Monte-Carlo model,
a GEANT simulation of the test beam measurements of the prototype modules with
$1\ \mathrm{mm}$ diameter KURARAY Y11(200)-M-DC fibers and 9903B was performed. 
Simulation was carried out both for realistic beam, including positron momentum spread 
in beamline and energy
loss upstream calorimeter, and for ideal beam.
 Comparison of simulation
with experimental measurements is displayed in Table \ref{tabl6}. Experimental 
and Monte-Carlo with realistic beam results match each other with an 
accuracy of a few percent. This gives us a confidence that Monte-Carlo simulation 
(with an ideal beam) properly reproduces the actual calorimeter resolution.

\begin{table}
\caption{Experimental and Monte-Carlo mean amplitudes and energy resolutions
for prototype Shashlyk module (240 layers of 1.5 mm scintillator and 0.35 mm lead,
$1\ \mathrm{mm}$ diameter KURARAY Y11(200)-M-DC fibers, 9903B). 
Amplitudes are normalized to the 
$1\ \mathrm{GeV}$ amplitude. Experimental energy resolution includes 
a contribution from beam momentum spread, energy loss upstream the calorimeter,
and effective noise.  Only statistical errors are displayed.}
\label{tabl6}
\begin{tabular}{ccccccc}
\hline
Beam          & \multicolumn{2}{c}{Measurements} 
              & \multicolumn{2}{c}{Monte-Carlo}   
           & \multicolumn{2}{c}{Monte-Carlo}   \\
momentum      & \multicolumn{2}{c}{ } 
              & \multicolumn{2}{c}{(realistic beam)}   
           & \multicolumn{2}{c}{(ideal beam)    }   \\
\cline{2-7}
$\mathrm{(GeV/c)}$       & $\langle A\rangle$ & $\sigma_E/E$   
              & $\langle A\rangle$ & $\sigma_E/E$   
              & $\langle A\rangle$ & $\sigma_E/E$   \\
\hline
0.5  & 0.514 & $6.29\pm0.17$ & 0.487 & $5.77\pm0.13$ & 0.507 & $5.17\pm0.08$ \\
1.0  & 1.\phantom{000} & $4.20\pm0.07$ & 1.\phantom{000} & $4.22\pm0.08$ & 
                                         1.017 & $3.84\pm0.06$ \\
1.5  & 1.509 & $3.30\pm0.05$ & 1.514 & $3.46\pm0.06$ & 1.526 & $3.24\pm0.05$ \\
2.0  & 2.043 & $2.92\pm0.05$ & 2.031 & $3.06\pm0.09$ & 2.037 & $2.88\pm0.05$ \\
\hline
\end{tabular}
\end{table}

     To understand the contributions of different effects on
the energy resolution of the module, a Monte-Carlo simulation of 
the signals produced by $250\ \mathrm{MeV}$ photons,  has been performed.  
The resulting energy resolutions  are displayed 
in Table \ref{Tab3}. 
Sampling effects were simulated in a very 
long (6000 layers) module.  Additional effects were consequently included.
Statistical accuracy of the calculation is 1.5\%.  
Some disagreement between calculations and test beam measurements may be related to the conditions of measurement (beam in the center of the module)
in which the effect of the nonuniformity of light collection has been suppressed.

\begin{table}
\caption{Calculated energy resolution 
 of prototype module (240 layers of 1.5 mm scintillator and 0.35 mm lead,
$1\ \mathrm{mm}$ diameter KURARAY Y11(200)-M-DC fibers, 9903B). 
uniformly exposed by  $250\ \mathrm{MeV}$ photons
depending on the level of detail of Monte-Carlo simulation} 
\label{Tab3}
\begin{tabular}{lc}
\hline
 & $\sigma_E/\sqrt{E\mathrm{(GeV)}}$ (\%) \\
\hline
  Sampling only                                        & 2.69  \\
  + 240 layers                                         & 2.84  \\
  + Holes and steel strips                             & 3.04  \\
  + Attenuation in fiber                               & 3.01  \\
  + Photostatistics                                    & 3.65  \\
  + Nonuniformity of light collection                  & 4.07  \\
  + $3\times3$ modules                                 & 4.27  \\
\hline
\end{tabular}
\end{table}

\section{Improving the energy resolution}

One can see that sampling adds the main contribution to energy resolution.
However, it is less than the combined contribution of other factors $3.3\%$
(adding in quadrature). Among these contributions the most significant 
are photostatistics $2.1\%$ and nonuniformity of light collection $1.8\%$. 
To reach the proposed resolution of $3\%/\sqrt{E\mathrm{(GeV)}}$ sampling, 
photostatistics and
uniformity of light collection must be improved.

The possibilities of improving the  sampling contribution are limited.
Decreasing the thickness of lead plates or increasing of thickness of 
scintillator plates increases the effective radiation length of the module, 
and as a result resolution deteriorates due to longitudinal fluctuations of 
electromagnetic shower and due to an increased transverse leakage 
(increased Moli{\` e}re
radius).
Simultaneously decreasing the thickness of lead and scintillator plates
will lower photostatistics.  Decreasing lead and/or scintillator
plates may also cause technical problems for module production.

\begin{figure}
\mbox{\includegraphics*[bb=0 15 270 270,width=\figwidth]{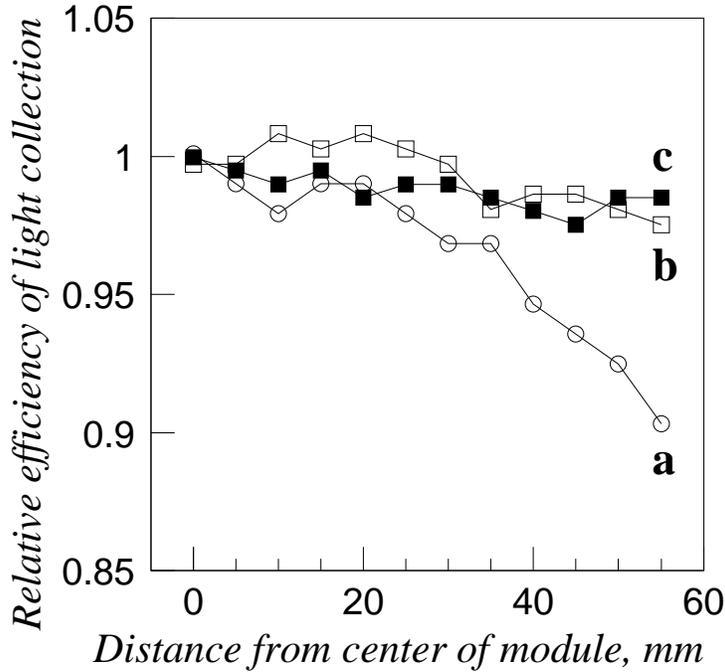}}
\caption{Monte-Carlo dependence of light output on distance from center of 
module for (a) existing prototype of module, (b) suggested module with scintillator plates totally 
covered by CMSS, and (c) for module with scintillator plates covered with CMSS only on side edge.
Distribution are normalized to 1 at center of plate.}
\label{FIG13}
\end{figure}

To improve the uniformity of light collection we consider using a 
chemical modification of a scintillator surface (CMSS) \cite{Semenov} 
on the edges of the scintillator plates, which produces thin (50-100 $\mu$m) 
white foam layer
with a diffuse reflection efficiency of about 93\%.
This coating also increase total light output (photostatistics).
Monte-Carlo distributions of light output 
as a function of the distance from the center of modules with and without this 
coating are shown in Fig. \ref{FIG13}.

Alternatively, uniformity of light collection may be improved by the appropriate
varying of the density of fiber location in the scintillator plate. 

In addition to increasing scintillator plate thickness and fiber diameter,
light output may be also increased by establishing optical contact between 
fiber and scintillator.  This may be accomplished by gluing fibers within 
scintillator holes. 
Direct measurements on a single plate (Fig. \ref{FIG11}) confirmed this 
concept.  The technology of producing this optical contact is
not yet developed for a total calorimeter, however, so calculations are made 
for both 
glue and air contact between fiber and scintillator. The gain in 
photostatistics with glue is expected to be about 2.

Within the limitations on sampling improvement we consider 3 possible 
versions of calorimeter. 
Results of calculations for the all versions of modules are presented in
Table \ref{Tab4}.

\begin{table}
\caption{Energy resolution ($\sigma_E/\sqrt{E\mathrm{(GeV)}}$) in a module, 
 uniformly exposed by $250\ \mathrm{MeV}$ photons,
depending on the level of detail of the Monte-Carlo simulation. Statistical 
accuracy of the calculations is 2\% } 
\label{Tab4}
\begin{tabular}{lccc}
\hline
 Module version  & 1 & 2 & 3 \\ 
\hline
Number of layers    & 400 & 300 & 300 \\
Lead thickness (mm) & 0.25 & 0.35 & 0.25 \\
Scint. thickness (mm) & 1.5 & 2.0 & 2.0 \\
Fiber diameter (mm)   & 1.2 & 1.2 & 1.2 \\ 
\hline
  Sampling only                      & 2.14 & 2.50 & 1.91 \\
  + Finite number of layers          & 2.12 & 2.54 & 2.13 \\
  + Holes and steel strips           & 2.34 & 2.68 & 2.31 \\
  + Attenuation in fiber             & 2.53 & 2.88 & 2.57 \\
  + Photostatistics                  & 2.82 & 3.09 & 2.75 \\
  + Nonuniformity of light collection   & 2.87 & 3.15 & 2.84 \\
\hline
  + $3\times3$ modules               & 3.17 & 3.42 & 3.23 \\
 $-$ gluing of fiber                 & 3.28 & 3.53 & 3.16 \\
\hline
\end{tabular}
\end{table}

One can see that improvements in mechanical and optical construction of modules
can yield energy resolution of about $3\%/\sqrt{E\mathrm{(GeV)}}$, and that modules with 
thinner lead plates might give better results.  There is almost no difference 
between the expected resolution for version 1 and 3.
A bigger longitudinal leakage fluctuations for version 3 are compensated
by better sampling term.

Resolution is approximately $0.2\%/\sqrt{E\mathrm{(GeV)}}$ worse if the shower is measured
by $3\times3$ modules. Increasing of the number of modules to capture a 
larger fraction of visible energy is limited by a greater noise contribution.

If fibers are not glued, resolution will be additionally 
increased by $0.1\%/\sqrt{E\mathrm{(GeV)}}$

\section{Conclusion}

\begin{figure}
\mbox{\includegraphics*[bb=0 15 270 270,width=\figwidth]{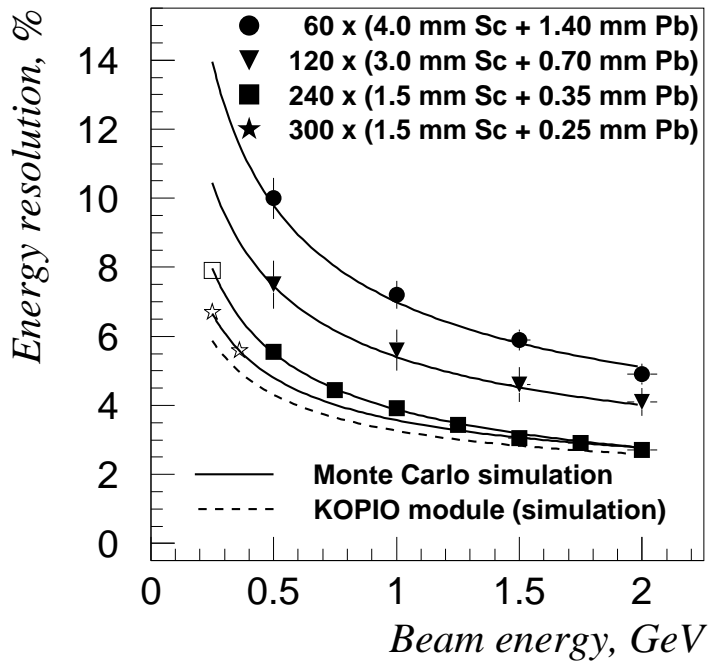}}
\caption{Comparison of the Monte-Carlo calculations (solid lines) of the energy resolution with our experimental measurements of Shashlyk modules,
({$\bullet$}) for BNL E865;
({\scriptsize$\blacktriangledown$}) prototype for BNL E923;
({\tiny$\blacksquare$}) prototype for KOPIO;
($\star$) other prototype for KOPIO (not described in this paper).
Sampling characteristic, number of layers and scintillator/lead thickness 
is displayed. Filled and empty marks corresponds to the measurements with positron and photon beam, respectively.
Dashed line indicates the expected energy resolution of the upgraded module, satisfying requirements of KOPIO experiment.
}
\label{FIG14}
\end{figure}

Modules for a Shashlyk calorimeter with energy resolution about $4\%/\sqrt{E\mathrm{(GeV)}}$
have been constructed and experimentally tested. 
Using these results as well as results of special measurements, 
the detailed Monte-Carlo simulation of the Shashlyk modules has been developed.
Our model of the Shashlyk module includes (i) the GEANT 3 simulation of an electromagnetic shower; (ii) the optical model for the simulation of light collection in the scintillator plates; (iii) effects of light attenuation in the fibers;
(iv) effects of quantum efficiency of the photodetectors, and electronic noise and thresholds in readout system. This model gives an excellent description of the energy resolution for variety of experimentally studied Shashlyk modules 
 which is illustrated in Fig. \ref{FIG14}.
A few measurements made with low energy photons \cite{photons} are also in perfect agreement 
with the developed model.

Our Monte-Carlo simulation indicates the possibility of improving this
resolution to about $3\%/\sqrt{E\mathrm{(GeV)}}$ for module with lead plate thickness 
0.25 mm. In a case with a realistic experimental 
environment, electronic noise and thresholds, and limited number of modules 
in a clump,  the realized resolution of 
such a calorimeter is expected to be about $(3.2\div3.3)\%/\sqrt{E\mathrm{(GeV)}}$.

     The experimental study of the test modules was performed
in a $0.5\div2.0~\mathrm{GeV/c}$ positron beam without precise measurement of 
the beam momentum.
For this reason extrapolation of the results of the measurements to the KOPIO 
energy region ($\sim 250\ \mathrm{MeV}$) is somewhat uncertain.  
The possible way to eliminate this ambiguity is a new test measurement
in a low momentum electron beam ($200\div500~\mathrm{MeV/c}$) with a controlled spread in beam momentum or a measurement with low energy photons 
\cite{photons}.

For future measurements we are planning to use the improved version of the KOPIO Shashlyk modules with expected energy resolution $(3.0\div3.5)\%/\sqrt{E\mathrm{(GeV)}}$.
These measurements will include a study of the APD photodetectors 
and Wave Form Digitizers for readout.

Experimental study of time resolution of Shashlyk module is also supposed to
be done in this measurements. 


\end{document}